\documentclass[a4paper]{jpconf}
\usepackage{graphicx}
\usepackage{amsmath}
\usepackage{type1cm}
\usepackage{float}

\usepackage{xspace}
\xspaceaddexceptions{]}

\newcommand{\dg}{\ensuremath{^\circ}\xspace}

\newcommand{\al}{\ensuremath{\alpha}\xspace}

\newcommand{\z}{\ensuremath{\zeta}\xspace}

\newcommand{\az}{\ensuremath{(\alpha,\zeta)}\xspace}

\newcommand{\fermi}{\textit{Fermi}~LAT\xspace}

\newcommand{\rtg}{radio-to-\ensuremath{\gamma} phase lag\xspace}

\newcommand{\gr}{\ensuremath{\gamma}-ray\xspace}

\newcommand{\figr}[1]{Fig.~\ref{fig:#1}\xspace}
\newcommand{\sect}[1]{Sec.~\ref{sec:#1}\xspace}

\newcommand{\eqn}[1]{Eq.~(\ref{eq:#1})\xspace}

\begin{document}
\title{Implementation of a goodness-of-fit test for finding optimal concurrent radio and \gr pulsar light curves}

\author{A~S Seyffert$^1$, C Venter$^1$, A~K Harding$^2$, J Allison$^3$ and W~D Schutte$^3$}
\address{$^1$~Centre for Space Research, North-West University, 11 Hoffman Street, Potchefstroom, 2531, South Africa}
\address{$^2$~Astrophysics Science Division, NASA Goddard Space Flight Center, Greenbelt, MD 20771, USA}
\address{$^3$~School for Computer, Statistical, and Mathematical Sciences, North-West University, 11 Hoffman Street, Potchefstroom, 2531, South Africa}
\ead{20126999@nwu.ac.za}

\begin{abstract}
    Since the launch of the {\it Fermi} Large Area Telescope in 2008 the number of known \gr pulsars has increased immensely to over 200, many of which are also visible in the radio and X-ray bands. Seyffert et al.\,(2011) demonstrated how constraints on the viewing geometries of some of these pulsars could be obtained by comparing their observed radio and \gr light curves by eye to light curves from geometric models. While these constraints compare reasonably well with those yielded by more rigorous single-wavelength approaches, they are still a somewhat subjective representation of how well the models reproduce the observed radio and \gr light curves. Constructing a more rigorous approach is, however, made difficult by the large uncertainties associated with the \gr light curves as compared to those associated with the radio light curves. Naively applying a $\chi^2$-like goodness-of-fit test to both bands invariably results in constraints dictated by the radio light curves. A number of approaches have been proposed to address this issue. In this paper we investigate these approaches and evaluate the results they yield. Based on what we learn, we implement our own version of a goodness-of-fit test, which we then use to investigate the behaviour of the geometric models in multi-dimensional phase space.
\end{abstract}

\section{Introduction}
    The \textit{Fermi} Large Area Telescope (LAT) is the primary instrument on the \textit{Fermi Gamma-ray Space Telescope} mission which was launched on 11 June 2008. To date \fermi has detected more than 200 new \gr pulsars, a dramatic increase over the 6 or 7 known at the time of \fermi's launch \cite{Thompson01}. This sudden abundance of pulsars visible in both the radio and \gr bands has opened the door to multiwavelength investigations that were not possible before.

    These studies have, however, been made difficult by the comparatively low \gr photon flux (typically), as compared to the radio photon flux for any particular pulsar. This disparity manifests itself in the relative uncertainties on the observed intensities in the two wavebands, making traditional goodness-of-fit techniques very ineffective when trying to fit the radio and \gr light curves (LCs) simultaneously. This has led some investigators to adopt so-called by-eye methods to find optimal simultaneous fits (e.g., \cite{Venter_MSP09,Seyffert2011}). A number of attempts have, however, been made to address this issue directly, and this paper presents our latest such attempt. The most notable attempts to date (\cite{Johnson,Pierbattista}) artificially inflate the uncertainties accompanying the radio LCs in an attempt to grant the radio and \gr data equal weight in the determination of the optimal fit. As we will show, neither of their approaches accomplish this fully. We will then describe our proposed solution using these two approaches as reference.

\section{The geometric pulsar models}\label{sec:model}
    We use an idealized picture of the pulsar system, wherein the magnetic field has a retarded dipole structure \cite{Deutsch55}. The \gr emission originates in regions of the magnetosphere (referred to as `gaps') where the local charge density is sufficiently lower than the Goldreich-Julian charge density \cite{GJ69}. These gaps facilitate particle acceleration and consequently radiation. We assume that there are constant-emissivity emission layers embedded within the gaps, in the pulsar's corotating frame. The location and geometry of these emission layers determine the shape of the \gr LCs, and there exist multiple models describing their geometry. In this paper only the two-pole caustic (TPC, \cite{Dyks03}) \gr model will be examined. For the radio emission we use a simple empirical hollow cone model \cite{Story}. For a more detailed description of these models, see our previous work \cite{Seyffert2011}. 

\begin{figure}[!ht]
        \centering
        \includegraphics[width=0.44\textwidth]{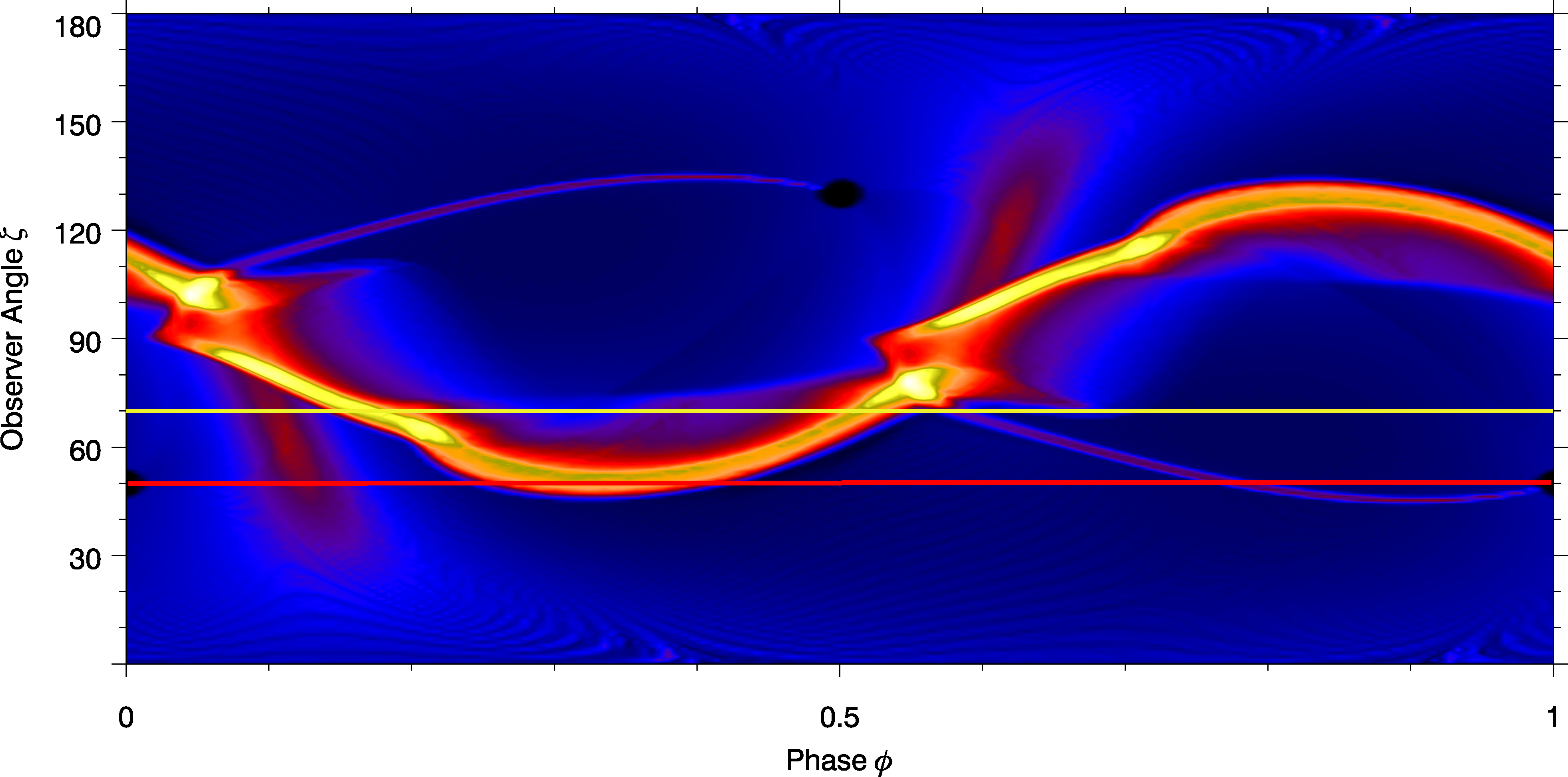}
        \includegraphics[width=0.22\textwidth]{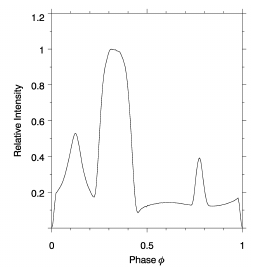}
        \includegraphics[width=0.22\textwidth]{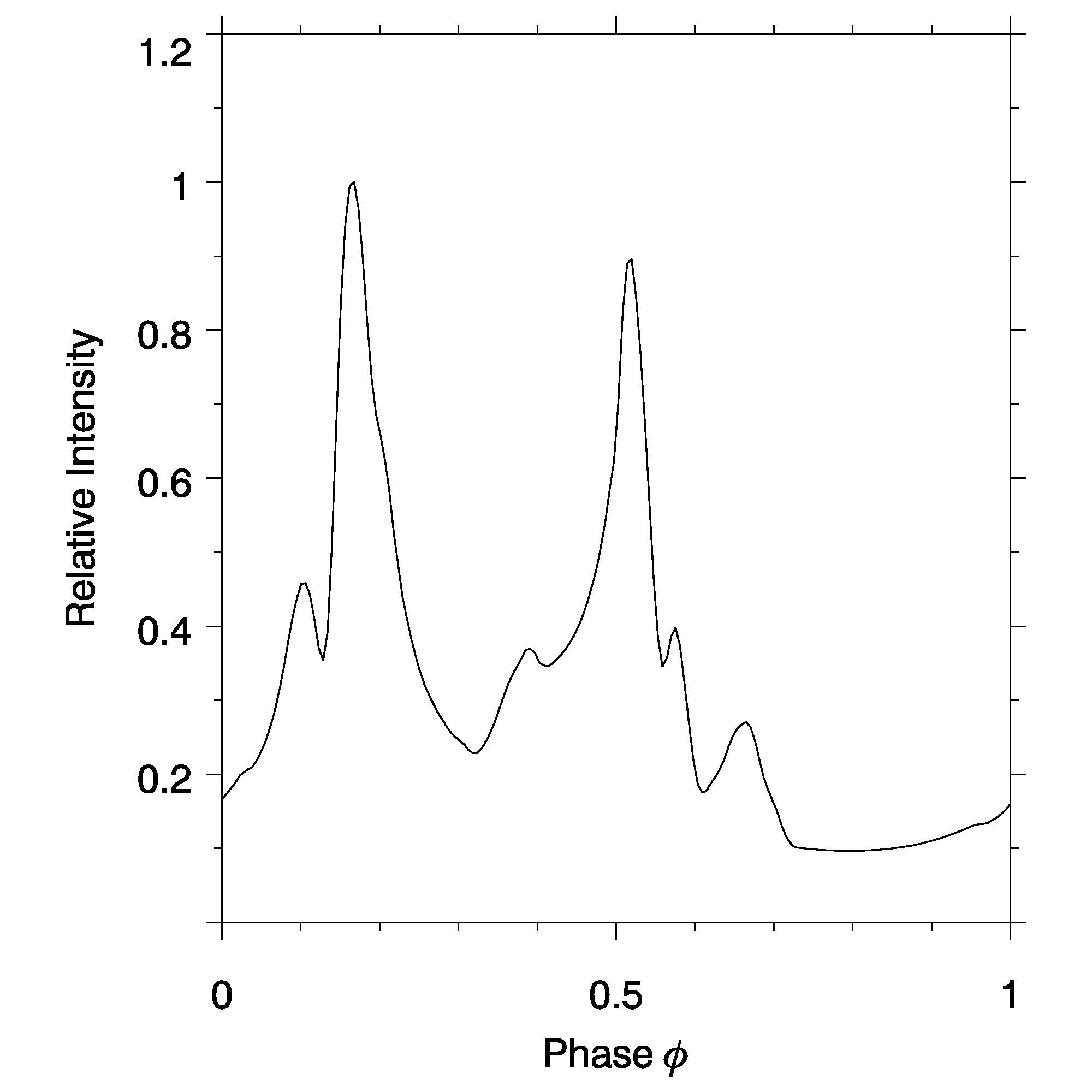}
        \caption{An example of constant-\z cuts through a phaseplot for $\al=50\dg$. The cuts correspond to values of $\z=50\dg$ (red) and $\z=70\dg$ (yellow). We see that the caustics on the phaseplot present as peaks in the LC. \label{fig:labeled}}
    \end{figure}

    \figr{labeled} (left) shows an example of a phaseplot produced using the TPC model. It is an equirectangular projection skymap of the emission of the pulsar, with rotational phase $\phi$ on the horizontal axis, and the observer's line of sight \z on the vertical axis. For each point in $(\phi,\z)$ space it gives the relative intensity per solid angle of the emission that would be observed if the observer's line of sight would pass through that point as the pulsar rotates. A phaseplot thus contains the projected all-sky \gr emission of a pulsar over one complete rotation.


    The equator of the neutron star is at $\z=90\dg$, and its magnetic poles are at $\z=\al$ and $\z=180\dg-\al$. Thus, each value of \al corresponds to a unique phaseplot. \figr{labeled} shows an example of how a predicted LC can be obtained by cutting a phaseplot at a constant value for \z (e.g., for a pulsar with $\al=50\dg$). The predicted LCs are independently normalised so their respective maxima are 1.

\section{The fitting methods}
    \subsection{The by-eye method}
        The by-eye method considers the LCs associated with the entire domain of \al and \z by generating so-called atlases. Each atlas consists of a set of predicted \gr and radio LCs generated at each point on a grid of \az values covering the \az domain at a chosen resolution. The data LCs are then compared to the model LCs at each point in this grid, and a subjective decision is made regarding the quality of each fit. Once the entire space has been scanned, a best-fit contour emerges, and constraints on the viewing geometry \az of the pulsar can be derived. An important aspect to note is that not only are the shapes of the LCs reproduced, but also the \rtg. For an example of this method's application, see our previous study \cite{Seyffert2011}.


        The by-eye method is not very rigorous, and would possibly yield different results when applied by different people. The results, however, do seem to be a good first attempt at finding concurrent fits when considered alongside other fits obtained independently (\cite{Weltevrede}, \cite{Seyffert2011}). Furthermore, this method allows us to discern which features of the LC most likely contribute to a specific LC's rejection, which in turn gives us insight into what effects are responsible for the shape of the final solution contours. Since the statistical methods aim to fit the same models (in our case at least) to the same data, we can use the knowledge garnered from the by-eye contours as a sanity check when considering the contours produced by the statistical methods.


    \subsection{The statistical methods}
        The statistical methods developed to date implement a modified version of Pearson's $\chi^2$ goodness-of-fit test, with the test statistic defined as

        \begin{equation}\label{eq:chiSq}
            X = \sum_{i=1}^{n_{\rm bins}} \left( \frac{E_i - O_i}{U_i} \right)^2,
        \end{equation}

        \noindent with $E_i$, $O_i$, and $U_i$ corresponding to the model (expected) intensity, observed intensity, and error (uncertainty) on the observed intensity for the $i$th bin (of $n_{\rm bins}$ bins) \cite{Avni}. This test statistic is $\chi^2$ distributed with $N=n_{\rm bins}-n_{\rm parameters}-1$ degrees of freedom. Ideally the minimum value obtained for $X$ (as a function of the tested model's parameters) will be roughly $N$, indicating that there is some set of parameters for which the model fits the data well. Since the current models are still only approximate representations of the actual phenomena, this unfortunately rarely occurs, with the best fits having values for $X_{\rm min}>>N$. To compensate for this, and to allow constraints to be derived from our imperfect models, the value of $X$ is normalised such that its minimum is $N$.

        This method has been applied successfully for LCs from a single waveband, but has proven to be less useful when trying to fit observed LCs from multiple wavebands concurrently. Calculating the values of $X$ for the LCs in each waveband and naively combining them (with equal weighting) for each set of parameter values, yields constraints that tend to be dominated by the waveband that has lower relative uncertainties. In our case, the radio goodness-of-fit test statistic dominates the overall fit.

\begin{figure}
            \centering
            \includegraphics[width=0.7\textwidth]{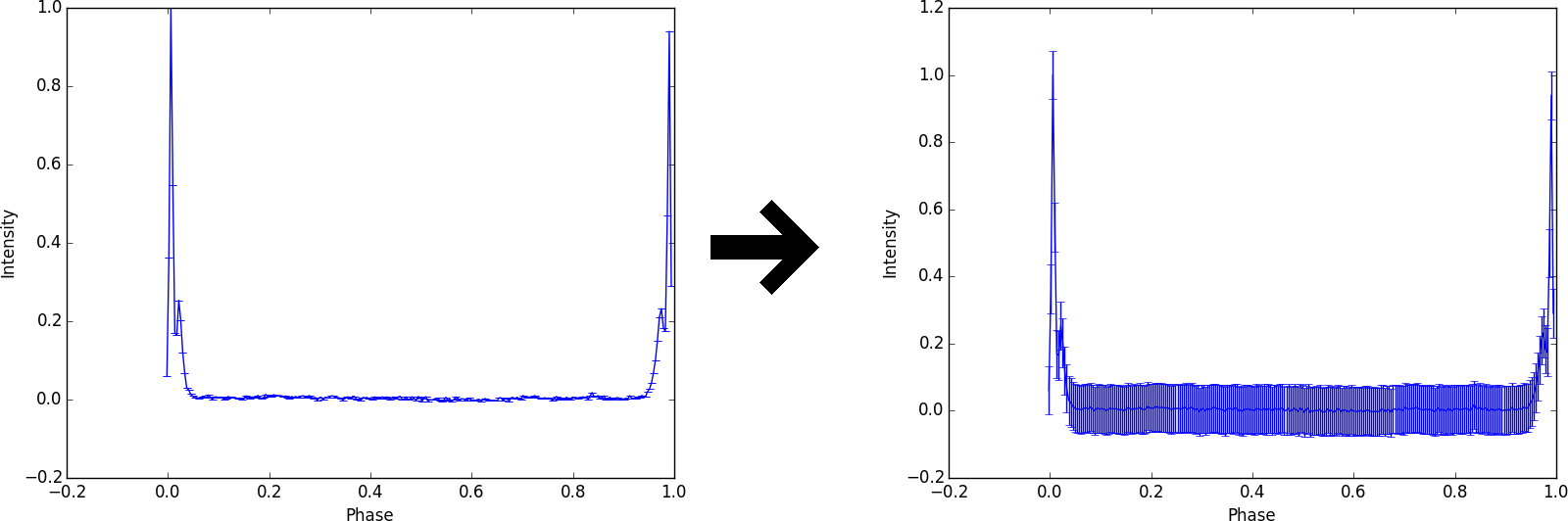}
            \includegraphics[width=0.7\textwidth]{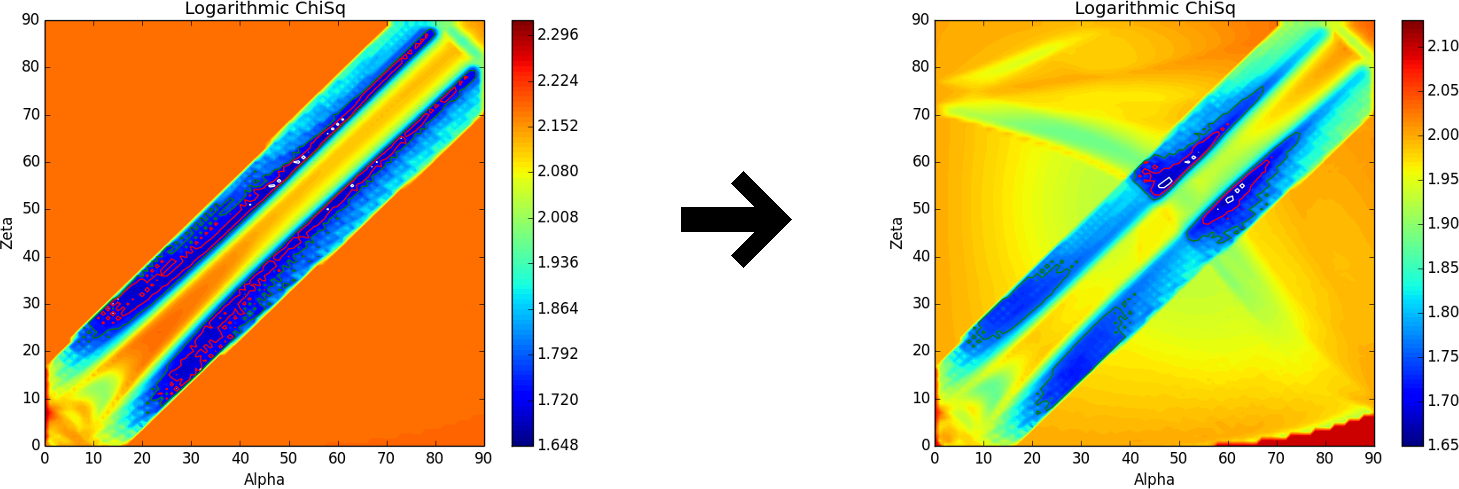}
            \caption{The top set of figures shows an example of radio uncertainties artificially inflated in a manner similar to that proposed by \cite{Johnson}. The bottom set of figures shows how this influences the resulting values for $X$ for the combined radio and \gr fits. The left-most of the $\chi^2$ maps finds its best fits along the diagonal, which means that the radio fit dominates. In the right-most $\chi^2$ map we see the cross-diagonal features of the \gr fit emerging. The resulting contours become more constraining as both models start contributing more equally to the fitting process. \label{fig:Johnson}}
        \end{figure}

        This problem has been addressed in the methods developed thus far, but with limited success. One approach is to somehow artificially inflate the radio uncertainties (e.g., \cite{Johnson,Pierbattista}). How this is done differs depending on the method, but the results are invariably an improvement on a naive goodness-of-fit method. \figr{Johnson} shows how these artificially inflated uncertainties look (specifically the method used by \cite{Johnson}), and how the resulting values of $X$ change.

        The motivation behind these statistical methods is that for us to be able to add the $\chi^2$ profiles for the radio and \gr goodness-of-fit tests, the values for $X$ associated with the best radio and best \gr fit should be of roughly comparable size. This is best accomplished by adjusting the uncertainties of the radio LC. These methods do not work effortlessly for all pulsars, though. For some pulsars, one needs to increase the degree to which the radio uncertainties are inflated, while for others a decrease is required \cite{Johnson}. This seems to indicate that there might still be room for improvement.

\section{Our proposed improvement}\label{sec:ourMethod}
    The method we propose takes a more direct route toward obtaining comparable radio and \gr values for $X$. In considering the statistical methods above we realised that ensuring comparable minima for the $\chi^2$ maps is not sufficient as the minima will generally not be colocated in \az space. Stated more clearly, we realised that the minimum of the \gr test statistic should be compatible with all the values of the radio test statistic, not just its minimum, and vice versa. This ensures that the minimum \gr test statistic value is not averaged with a (non-minimum) radio test statistic value that is orders of magnitude larger, when combining the radio and \gr $\chi^2$ maps, since the radio will still dominate in that case.


    To accomplish this compatibility we scale the radio test statistic values such that their dynamic range (the difference between their maximum and minimum) is equal to that of the \gr test statistic values. In effect we force the radio and \gr test statistic values to carry equal weight in the determination of the best concurrent fit. This scaling is done globally over the space defined by the parameters for which we are deriving constraints.

    The procedure employed proceeds as follows:

    \begin{enumerate}
        \item \textbf{Calculate the unscaled $\chi^2$ maps, using \eqn{chiSq}, for the radio and \gr fits independently.} In our case each model LC has four parameters associated with it, which means these maps will have four dimensions. The LC parameters are \al, \z, $\phi_0$ (zero point in phase), and $A$ (the LC amplitude).
        \item \textbf{Minimise each map over all parameters that are independent for the two wavebands.} For the concurrent fits in this study only $A$ can be minimised over in this step. Since we assume that the radio and \gr emission originates from different regions in the same magnetic field, the value of $\phi_0$ cannot be varied independently, and can therefore not be minimised over in this step. The resulting maps in our case have three dimensions.
        \item \textbf{Scale the \gr map to have a minimum equal to the appropriate number of degrees of freedom \cite{Pierbattista}.} For simplicity, we assume here that this scaled test statistic is $\chi^2$ distributed.
        \item \textbf{Scale the radio map such that it has the same dynamic range as the freshly-scaled \gr map.} The scaled radio test statistic is also assumed to be $\chi^2$ distributed, in accordance with the assumption made for the scaled \gr test statistic.
        \item \textbf{Sum the two maps.} The resulting combined map will have as many dimensions as the models have coupled parameters (parameters that cannot be varied independently). In our case the coupled parameters are \al, \z, and $\phi_0$. The number of degrees of freedom for this combined test statistic is the sum of the minima of the constituent maps \cite{Lampton}. This follows from the assumption made in \textit{(iii)} and \textit{(iv)} for the two constituent maps.
        \item \textbf{Minimise the combined map over all coupled parameters except \al and \z.} For our study, and more generally, these two parameters are the ones for which constraints are to be derived. The resulting map therefore has two dimensions.
        \item \textbf{Obtain the confidence contours using the $1\sigma$, $2\sigma$, and $3\sigma$ values of a $\chi^2$ distribution with 2 degrees of freedom.} Since the lowest value the minimum of the combined map can have is the sum of the minima of the constituent maps (in the case of colocated minima), we can derive the contours by adding the above-mentioned $1\sigma$, $2\sigma$, and $3\sigma$ values to this lowest minimum \cite{Lampton}.
    \end{enumerate}

\section{Results}
    Implementing this approach for the six pulsars we previously studied yields constraints that compare very well to those obtained by eye \cite{Seyffert2011}. The inclusion contours shown were derived as discussed in step \textit{(vii)} in the previous section. They should, however, only serve as an estimation of how the actual contours may look since it is not yet clear how the actual contours should be computed.

    \begin{figure}
        \centering
        \includegraphics[width=0.89\textwidth]{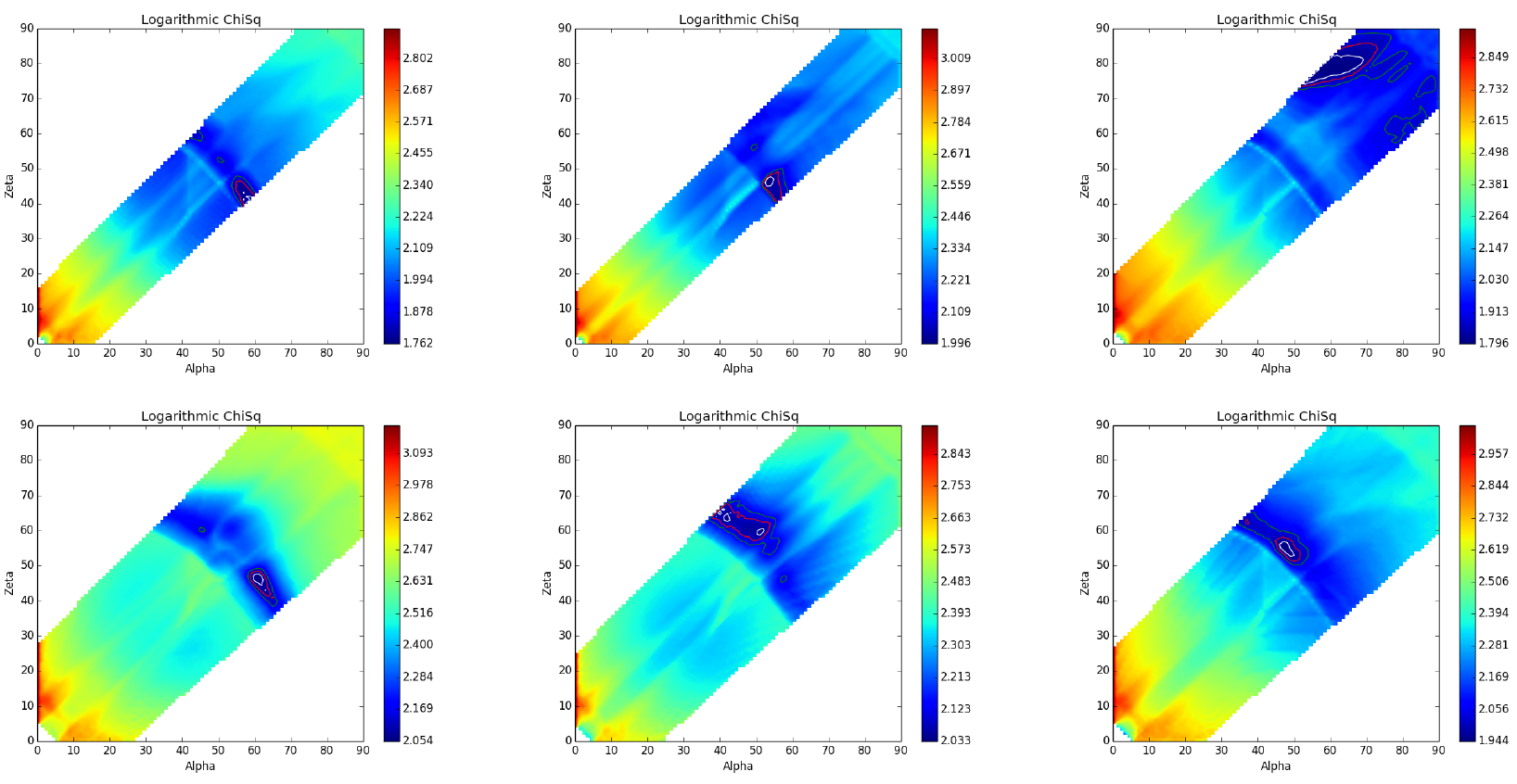}
        \caption{The resulting combined $\chi^2$ maps for the six pulsars studied in our previous work \cite{Seyffert2011}. The white, red, and green contour lines represent $1\sigma$-, $2\sigma$-, and $3\sigma$-like contours respectively, but should be considered cautiously. The white regions are where the radio model predicts no LC. \label{fig:newResults}}
    \end{figure}

\section{Conclusion and Future Work}
    The inclusion contours obtained using our proposed method compare favourably with those for the by-eye method, which in turn compare well with independently derived constraints \cite{Seyffert2011}. Although the the aim of these statistical methods is not to reproduce the results obtained using the by-eye method, the results of the by-eye method serve as a sanity check which the statistical methods should pass.

    The $\chi^2$ maps produced using our method have a number of properties which may prove useful in future studies. Chief among these properties is that the contours are calculated relative to a value other than the minimum of the combined map. This allows our method to reject an unsuccessful model combination even though some normalisation is done in obtaining the combined map. Another useful property is that the combined map remains essentially unchanged if step \textit{(iii)} is omitted, effectively only being scaled. This allows for the possibility of comparing the goodness of fit of different model combinations, and therefore identifying the most successful model combination.

    The interplay between our confidence in the geometric models and the uncertainties associated with the observed LCs also seems to be an important one to understand. It is useful to differentiate between evaluating which predicted LCs fit \textit{best}, and whether or not the observed LCs are fit \textit{well}. For a single model, determining the goodness of fit for each possible solution not only tells us how \textit{well} the data is fit, but also allows us to determine which of those solutions is the \textit{best} fit. This correlation does not hold when we generalise our method to include multiple models and multiple datasets. The breakdown occurs because we calculate the combined test statistic as the sum of the two constituent test statistics, making it sensitive to changes in the scaling of either of the single-model test statistics. Since our confidence in each model could be expressed as a scaling of the test statistic, effectively weighting their contribution to the combined test statistic, it is difficult to ensure that the single-model test statistics are properly scaled before summation. Our approach to the scaling of the individual test statistics is an attempt to address this problem directly.


    The most important open issue in this study is how the confidence contours are to be defined. The specification given in \sect{ourMethod}, based on the assumptions regarding the distributions associated with the radio and \gr test statistics, still needs to be tested. We are currently developing a Monte Carlo simulation strategy aimed at doing this. The method derived here will also be used to compare a variety of multi-model combinations based on concurrent fitting of pulsar LCs.

\ack This work is based on research supported by the National Research Foundation (NRF) of South Africa (Grant Numbers 90822 and 93278). Any opinions, findings, and conclusions or recommendations expressed are that of the authors, and the NRF accepts no liability whatsoever in this regard. A.K.H. acknowledges support from the NASA Astrophysics Theory Program. C.V. and A.K.H. acknowledge support from the \textit{Fermi} Guest Investigator Program.

%

\end{document}